\documentclass[12pt]{spieman}  
\usepackage{amsmath,amsfonts,amssymb}
\usepackage{graphicx}
\usepackage{setspace}
\usepackage{tocloft}
\usepackage{textcomp}
\usepackage[T1]{fontenc}
\usepackage{dirtytalk}
\usepackage{lineno}

\title{Reduced motion artifacts and speed improvements in enhanced line-scanning fiber bundle endomicroscopy}

\author[a,*]{Andrew D. Thrapp}
\author[a]{Michael R. Hughes}
\affil[a]{Applied Optics Group, School of Physical Sciences, University of Kent, Ingram Building, Canterbury, CT2 7NH, United Kingdom}

\cftpagenumbersoff{figure}
\cftpagenumbersoff{table} 
\begin{document} 
\maketitle

\begin{abstract}

\textbf{Significance:} Confocal laser scanning enables optical sectioning in fiber bundle endomicroscopy but limits the frame rate. To be able to better explore tissue morphology it is useful to stitch sequentially acquired frames into a mosaic. However, low frame rates limit the maximum probe translation speed. Line-scanning confocal endomicroscopy provides higher frame rates, but residual out-of-focus light degrades images. Subtraction based approaches can suppress this residue at the expense of introducing motion artifacts.

\textbf{Aim:} To generate high frame rate endomicroscopy images with improved optical sectioning, we develop a high-speed subtraction method that only requires the acquisition of a single camera frame.

\textbf{Approach:} The rolling shutter of a CMOS camera acts as both the aligned and offset detector slits required for subtraction-based sectioning enhancement. Two images of the bundle are formed on different regions of the camera, allowing both images to be acquired simultaneously.

\textbf{Results:} We confirm improved optical sectioning compared to conventional line-scanning, particularly far from focus, and show that motion artifacts are not introduced. We demonstrate high-speed mosaicing at frame rates of up to 240 Hz.

\textbf{Conclusion:} High-speed acquisition of optically sectioned images using the new subtraction based approach leads to improved mosaicing at high frame rates.
  
\end{abstract}


{\noindent \footnotesize\textbf{*}Andrew Thrapp,  \linkable{at600@kent.ac.uk} }

\begin{spacing}{2}   

\section{Introduction}
\label{sect:intro}  

A conventional biopsy involves a clinician extracting suspect tissue from the body. The tissue is subsequently processed and examined in a pathology lab for morphological or cellular changes to determine whether a disease is present or stage its progression. Fluorescence endomicroscopy, a probe-based endoscopic technique with cellular-scale resolution, provides a means to perform a comparable analysis \textit{in vivo} and in real-time. Fluorescence imaging is sometimes feasible using the autofluorescence of endogenous tissue structures such as collagen, but more commonly requires application of topical or intravenous contrast agents \cite{RN112}. 

Endomicroscopy is often performed with a flexible probe that is guided through the working channel of an endoscope. Working channels are typically less than 4 mm in diameter, with some variation among manufacturers and clinical applications. Therefore, the endomicroscopy probes must have diameters of around 3.5~mm or smaller and a sufficiently short rigid tip to traverse the endoscope as it bends.

As with all \textit{in vivo} microscopy techniques, optical sectioning is beneficial in fluorescence endomicroscopy to reduce corruption of images by out-of-focus fluorescence. Clinical studies with intravenous fluorescein are generally conducted using optical sectioning endomicroscopes. While careful use of topical stains can allow non-sectioning endomicroscopes to resolve surface features successfully \cite{RN49, RN427} there is still a clear benefit to optical sectioning in some cases \cite{RN108}.

As a means of providing sectioning, confocal laser scanning microscopy has been adapted to endomicroscopy using both distal and proximal scanning architectures \cite{RN75, RN1}. Distally-scanned systems use a single-core fiber to act as both the point source and confocal detection pinhole. This approach requires precise miniaturized distal scanning elements to be integrated at the tip of the probe, and the only commercially available system developed and subsequently withdrawn (ISC-1000, Optiscan/Pentax) provided frame rates of only up to 2~Hz. \cite{RN165}. Proximal scanning architectures instead make use of a multicore (\textasciitilde 30,000 core) fiber-optic imaging bundle. The imaging bundle relays excitation light from a bulk proximal scanning system (such as galvanometer scanning mirrors) and collects the returning fluorescence emission before de-scanning \cite{RN75}. The currently commercially available endomicroscope using this approach (Cellvizio, Mauna Kea Technologies) achieves frame rates of up to 12~Hz \cite{RN112}. However, the resolution is limited by the spacing between the fiber cores (down to approximately 3~\textmu m), although this can be further improved at the expense of the field-of-view by using a distal objective with non-unity magnification.

The field-of-view of endomicroscopy probes is typically less than 1~mm, far smaller than a typical histology slice. It is possible to increase the effective image size by collecting a time-series of images as the probe is moved and then computationally stitching these images into a mosaic. Mosaicing requires the probe shift between image frames to be determined, which can be done in real-time using pairwise rigid registration \cite{RN82}. More computationally intensive approaches, which are typically done offline, can, for example, embed low-resolution images into a higher precision mesh and hence improve resolution \cite{RN76}, or correct for non-rigid distortions \cite{RN38}. 

In both real-time and offline mosaicing, the imaging frame rate determines the maximum speed at which the probe can be moved due to the need to ensure sufficient overlap between images for registration. For example, for a probe with a 500 \textmu m diameter field-of-view, and assuming that there can be a maximum shift of 1/3 of the image diameter between consecutive images, the maximum speed the probe can be translated at for acquisition frame rates of 2 Hz and 12 Hz are, respectively, 0.33 and 2.0~mm/s. These low translation speeds are difficult to achieve reliably and consistently in realistic clinical scenarios. If distal optics are used to improve resolution and thus decrease the field-of-view, the allowable speeds are reduced even further by a factor equal to the magnification. Therefore, high frame rates are an important factor determining the feasibility of mosaicing.\cite{RN37}

A higher frame-rate alternative to conventional confocal endomicroscopy - and the topic of this paper - is to use line-scanning illumination. Images are formed by scanning a laser line over the fiber bundle's proximal face, and the fluorescence emission returning from the sample is then imaged onto a detection slit and subsequently onto a camera \cite{RN138}. This technique, known as line-scanning or slit-scanning confocal microscopy, was first adapted to endomicroscopy by Sabharwal~et~al.\cite{RN152}. The method of Sabharwal~et~al.\cite{RN152} used a configuration where returning fluorescence was de-scanned and imaged through a physical detection slit before being re-scanned onto a 2D camera. Similar systems were used for several subsequent studies, including multispectral imaging \cite{RN428}. A faster configuration using a 1D linear CCD, and hence not requiring re-scanning, was later reported by Hughes~et~al.\cite{RN37}, achieving a frame rate of over 100~Hz.

An alternative to a physical detection slit or linear camera is to use the rolling shutter of a CMOS camera as an electronically-variable moving slit with a width that can be changed on-the-fly. The use of the electronic slit was first demonstrated for benchtop systems by Mei~et~al.\cite{RN429} and later adapted to endomicroscopy by Hughes~et~al.\cite{RN68} Providing that the camera readout is synchronized with the scanning laser line, this avoids the need for returning fluorescence to be de-scanned by the scanner; the bundle can be imaged directly onto the CMOS camera (via appropriate fluorescence filters). The frame rate is then limited only by the camera readout speed; Hughes et al. achieved 120 Hz \cite{RN68}.

While line-scanning provides a degree of optical sectioning, it is not as effective as conventional point-scanning confocal microscopy \cite{RN129}. While close to focus the performance is similar, far from focus the amount of out-of-focus light rejected only increases approximately linearly with the distance from focus rather than quadratically as in point-scanning confocal. Fluorescence from far-from-focus depths can, therefore, still significantly degrade the image. 

To mitigate the far-from-focus light degrading images, a two-frame subtraction based approach, previously shown in principle for benchtop systems, was adapted to endomicroscopy \cite{RN129, RN68}.  The first frame (the confocal image) of the sequence is acquired with the rolling shutter virtual detector slit aligned with the laser line as it is scanned, and the second (the residual image) is acquired with a fixed offset between the laser line and detector slit. The residual or offset image provides a first-order estimate of the residual out-of-focus signal not rejected by the slit. The residual image is then subtracted from the confocal image to generate an enhanced line scan (ELS) image with improved optical sectioning. The full theory demonstrating how the subtraction of the residual image leads to an image with optical sectioning properties comparable to point-scanning confocal, albeit with inferior signal-to-noise ration, is presented in Poher et al.\cite{RN129}.


Since two frames are required to synthesize each ELS image, this subtraction approach leads to a halving of the frame rate and, as with any multi-frame approach, introduces motion artifacts when the probe shifts between acquisitions. Motion artifacts present particular problems for mosaicing, where such shifts are required and deliberately introduced. A two-camera approach was recently proposed, which avoids these inter-frame motion artefacts\cite{RN101}. Rather than acquiring the two images sequentially, the beam is split between two rolling shutter cameras, one with triggering timing adjusted to provide an aligned-slit (the confocal image) and one providing an offset-slit (the residual image). The authors used an LED source to generate the excitation line, with the scanning line created by displaying a series of patterns on a digital micromirror device (DMD), a programmable array of mirrors. A DMD has the advantage of removing the need for a laser or scanning system but has several limitations, including very low throughput of power from the LED (since at any one time most of the light is blocked) and that the DMD can only store a limited number of patterns onboard, insufficient to generate a full set of scanned line positions. Multiple parallel scanning lines were used to overcome this, rather than a single swept line, allowing the same DMD pattern to be used for multiple scan positions. Parallel scanning has several drawbacks: the width of the illumination lines is constrained by the number of lines which can fit onto the DMD array, the size of the detection slit is limited by the spacing between the illumination lines, and it requires the excitation of regions where light is not being collected which may contribute to scattering and photo-bleaching. 

A two-camera approach with a galvanometer scanner used to scan a laser line is feasible but would require the inclusion and fine alignment of a second camera, and the need to trigger both cameras typically reduces the frame rate by a factor of two. Instead, we now report a single camera configuration of the ELS approach for a galvanometer based system which restores the full frame rate and eliminates motion artifacts. We denote this approach as ELS+. The system projects two images of the bundle onto different parts of a camera to introduce a controllable offset. Hence, using the camera's rolling shutter as both the aligned and offset detection slits, the confocal and residual images can be provided simultaneously. The system can achieve enhanced sectioning at frame rates of up to 240~Hz with a lateral resolution of 8.8~\textmu m. Since the two images are acquired simultaneously, the approach does not suffer from inter-frame motion artifacts. The optical sectioning performance is essentially identical to ELS using sequential/alternating acquisition of the two frames, which we denote ELSA. ELS+ provides a frame rate improvement over the previously reported single-camera approach\cite{RN68} (60 Hz) by a factor of four and over the two-camera DMD approach\cite{RN101} (15 Hz) by a factor of 16. We also demonstrate real-time image registration and mosaicing at the full frame rate, which, to our knowledge, is the fastest endomicroscopy mosaicing system reported to-date.

\section{Methods}
\begin{figure}
\centering\includegraphics[width=12cm]{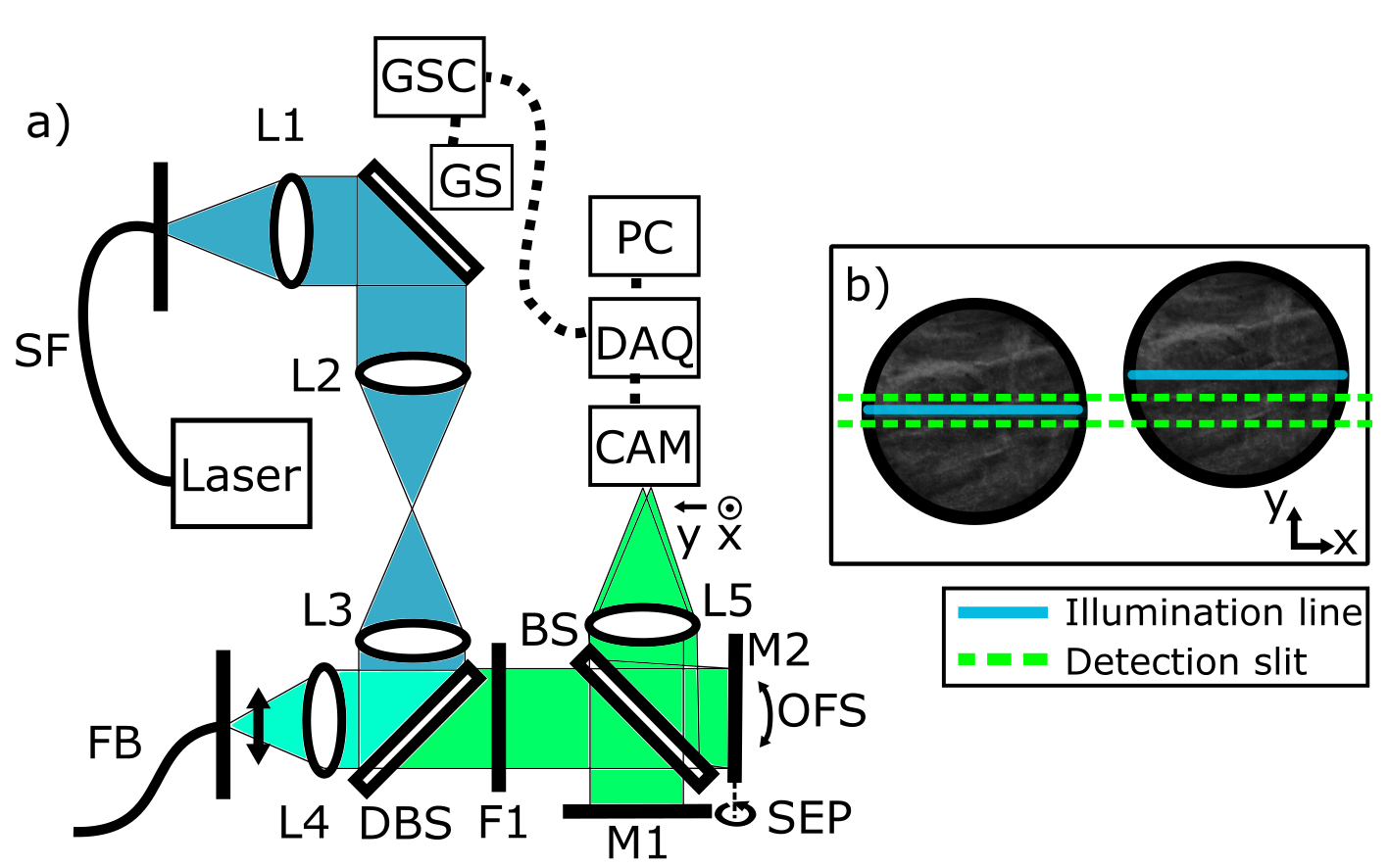}
\caption{(a) ELS+ system schematic: L1 - 4x finite conjugate objective, L2 - f = 50 mm cylindrical lens (power axis only is shown), L3 - f = 50 mm lens, L4 - 10x infinity-corrected objective, L5 - f = 40 mm tube lens, GS - galvanometer scanning mirror (galvo), GSC - galvanometer scanner controller,  DAQ - data acquisition card, SF - single mode fiber, FB - fiber bundle (scanning direction across bundle shown), DBS - dichroic beam-splitter, F1 - emission filter, BS - beam-splitter, M1 - stationary mirror, M2 - adjustable tilt mirror with image offset axis (OFS) and separation axis (SEP) shown. (b) Illustration of camera plane showing the offset residual image and an example of the positions of the scanning illumination line and detection slit due to the rolling shutter.}
\label{schematic}
\end{figure}

A schematic diagram of the ELS+ endomicroscopy system is shown in Fig. \ref{schematic}(a). The output from a fiber-coupled 488~nm laser diode (Thorlabs L488P60) is collimated by a 4x microscope objective and passes onto a single galvanometer (galvo) scanning mirror (Thorlabs GVS001), then onto the back aperture of a 10x infinity-corrected objective (Nikon PLAN N RMS10X) via a 50~mm focal length cylindrical lens (Thorlabs LJ169SRM-A), a 50~mm focal length achromatic doublet lens (Thorlabs AC254-050-A-ML) and a dichroic mirror (Thorlabs DMLP490). A laser line is produced on the face of a 30,000 core fiber optic bundle (Fujikura FIGH-30-650S), which can be scanned by the galvo mirror. 

The fiber probe does not use any distal optics, and so the probe is operated in direct contact with the tissue for imaging. The bundle's image diameter is approximately 620~\textmu m, and the average spacing between the bundle cores is 3.2~\textmu m. Fluorescence returning through the bundle is collected by the objective and transmitted by the dichroic before passing through an emission filter (Thorlabs DLMP490). It is then split into two paths by a beam-splitter onto a fixed mirror and a tilting mirror. The mirror configuration is designed to guide identical images of the bundle onto the left and right regions of a rolling shutter CMOS camera (Flea 3, FLIR FL3-U3-13S2M-CS) via a 75~mm tube lens (Thorlabs AC254-075-A-ML).

The rolling shutter camera uses sequential line readout, which progresses at a rate of 7.63~\textmu s per line. The exposure determines the slit width by controlling how many lines are exposed simultaneously. The magnification factor between the sample and the camera was 3.8. Therefore, the minimum slit width, determined by the dimensions of a camera pixel, was 3.63~\textmu m at the camera plane and 0.96~\textmu m at the bundle plane. The relationship between the number of simultaneous active lines, $N$, and exposure, $E$ in microseconds is 
\begin{equation}
    \mathrm{N} = \frac{\mathrm{E}}{7.63 \, \text{\textmu s},}
\end{equation}
where $N$ is the number of active lines. The slit width at the bundle plane, $W$, is therefore given by
\begin{equation}
    \mathrm{W} = \frac{\mathrm{M \Delta p E}}{7.63 \, \text{\textmu s},}
\end{equation}
where $M$ is the magnification factor, and $\Delta p$ is the pixel size.

The camera is operated in free-run mode, and the strobe signal from the camera is used to trigger the generation of each galvo scanning waveform signal via a USB DAQ Card (National Instruments). Synchronization between the ramp waveform driving the galvo and the camera readout is achieved via a calibration procedure described below.

Two images of a bundle are shown illustratively on the camera plane in Fig. \ref{schematic}(b). The offset of the bundle's right-hand image (the residual image) on the camera is controllable with the tilting mirror (M2), allowing the rolling shutter to act as both the aligned and offset detection slit with a manually adjustable offset.

The camera can be run at the standard frame rate (120~Hz) and at a faster rate (240~Hz), in which case only a smaller region of interest can be used. When running at 120~Hz, the setup described above gives 6.8 camera pixels per average fiber core spacing (high sampling), leading to a lateral resolution of 5.52 \textmu m, as measured using a back-illuminated USAF resolution target. To operate at 240~Hz, the 10x imaging objective was replaced with a 4x objective (finite conjugate). The smaller bundle image led to slight under-sampling of the bundle by the camera; the bundle was sampled by 2.4 camera pixels per fiber core spacing (low sampling), leading to a slightly reduced resolution (due to under-sampling) of 8.8 \textmu m. An alternative approach would have been to crop the bundle using a mask at conjugate plane, allowing the full resolution to be maintained at the cost of slight reduction in the field-of-view.

The power delivered to the sample was measured to be 1~mW. The losses introduced by the beam-splitter configuration alone were 50\%, and each image has a further 50\% drop when the intensity is divided between the two images. It would be possible to minimize the 50\% loss at the beam-splitter using a polarizing beam-splitter and quarter-wave plate configuration. Excitation arm losses before the bundle include reflections, filter absorption losses, and those associated with coupling to the bundle; these can easily be compensated by increasing the laser power.

\subsection{Image Processing}
The raw image consists of two images of the fiber bundle, the line-scan confocal image, on the left side of the camera, and the residual image, on the right. A background estimation is acquired by placing the probe in a dark tube and averaging 50 images.  The laser was on during background acquisition to account for auto-fluorescence generated in the fiber as well as room light and stray reflections in the optical system. The two images are then extracted, using a prior position calibration discussed in Sect.~\ref{Sect:SystCalib}, and the residual is subtracted from the confocal to generate the ELS+ image. The image is then smoothed with a 3x3 kernel median filter, similar to the method described by Pierce et al. \cite{RN78}, to eliminate the honeycomb fiber core structure, especially noticeable in the high sampling mode/low frame rate images. Images acquired with low sampling had inherent smoothing due to the under-sampling of the bundle pixels by the camera and so were not filtered further. 

Processed ELS+ images can be stitched together into a mosaic in real-time. The images are first down-sampled to 250x250 pixels and then circularly cropped to 90\% of the bundle diameter. The shift between each pair of successive images is then registered using normalized cross-correlation between a template extracted from one image and the entire second image \cite{RN426}. The current image is then inserted dead-leaf at the correct position in the mosaic, where the pixel values of the new frame completely overwrite any previous pixel values in the mosaic. This algorithm was implemented mostly in Labview (National Instruments, Austin, TX) while the normalized cross-correlation was performed using a DLL written in C++ calling OpenCV libraries. The processing frame rate was sufficient to allow mosaicing at the maximum frame rate (240~Hz) on a PC with an Intel i7-7700 Quad-Core processor and 8~GB RAM. However, with the exception of the demonstration of real-time mosaicing at 240~Hz, mosaics shown in this article were constructed off-line from saved stacks of images to allow for detailed analysis and archiving of raw data.

\subsection{System Calibration}
\label{Sect:SystCalib}
The system is calibrated using a custom-built LabVIEW VI. The process consists of three steps: (i) setting the slit-offset for the residual image to the desired value, (ii) synchronizing the laser line-scanning with the camera readout by mapping the galvanometer drive signal voltage to the location of the line on the camera (in pixels), and (iii) precisely locating the bundle images on the CMOS so that subtraction can be performed. These calibration steps only need to be performed once for a given setup.

\begin{figure}
    \centering
    \includegraphics[width=13cm]{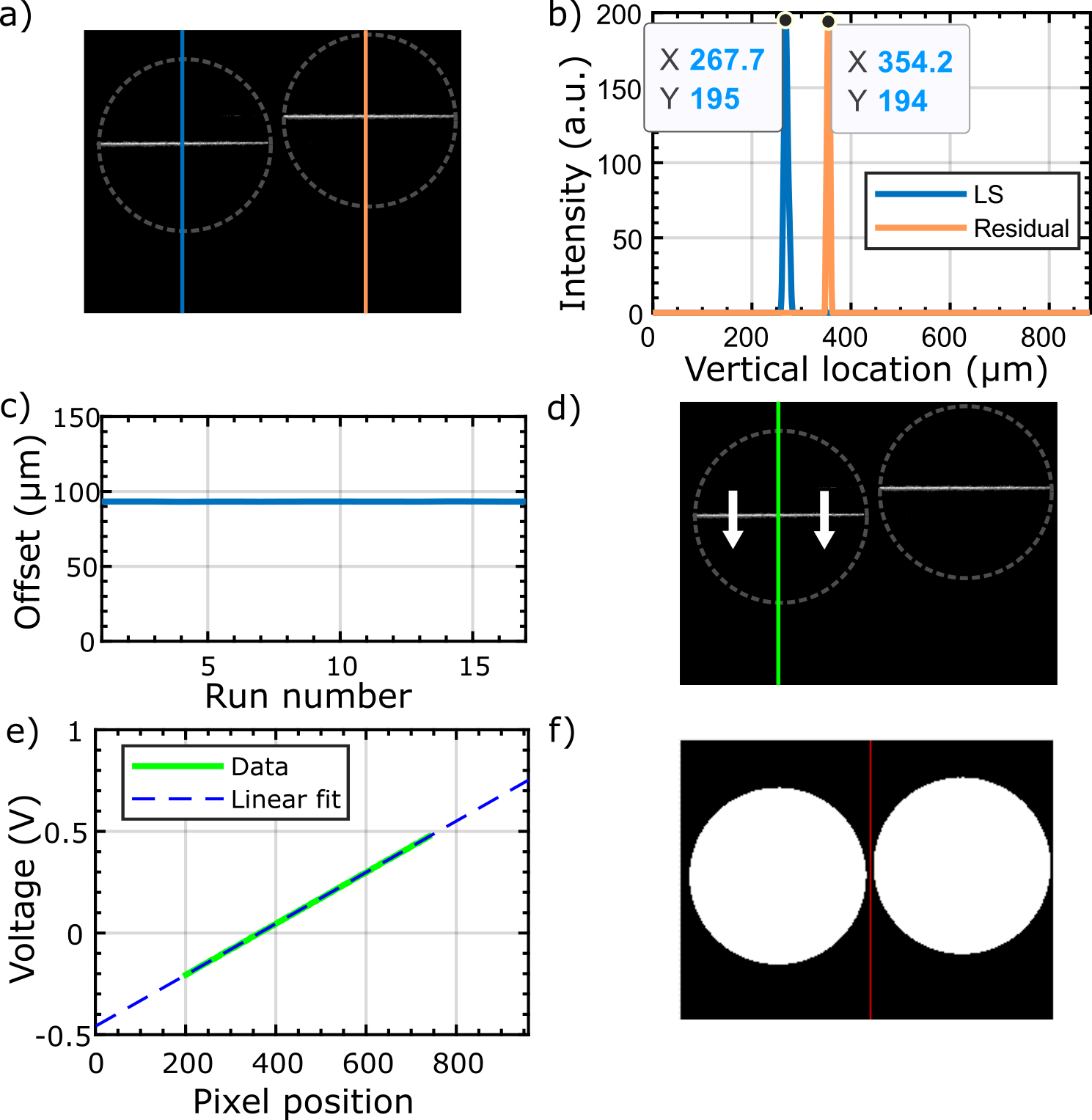}
    \caption{System calibration. (a-c) Calibration of line offset. (a) Non-scanning laser line visible on the bundle for the aligned image (left) and offset image (right). (b) Intensity profiles were taken perpendicular to the laser lines for the two images. (c) A difference of the central values determined by Gaussian fits of the two intensity profiles (the offset) reported in real-time. (d,e) Mapping of galvo voltage to pixels on CMOS camera. The line is slowly swept across the bundle face in the direction of the arrow, and (d) at each position step, an intensity profile is generated across the line. (e) The pixel position with the highest intensity is reported for each voltage, and a line is fit with the slope and intercept, determining the mapping between galvo voltage and line position in pixels. A typical mapping is shown, slope~=~ 0.00126~V/pixel, offset~=~-0.459~V, $R^{2} = 0.99$. (f) Bundle image locator routine. The bundle is illuminated from the distal end with an LED, and a binary threshold filter is applied. The centers of the two bundle images are returned as the rows and columns of the bundle with the greatest number of pixels above the threshold.}
    \label{fig:Calibration}
\end{figure}

The first step is to set the offset of the residual image from the confocal image so as to achieve the desired optical sectioning performance (which, as discussed later, must be balanced with signal-to-noise requirements). The offset calibration procedure is illustrated in Fig. \ref{fig:Calibration}(a-c). Since the offset between the images should be perpendicular to the laser line (`vertical' direction as shown in Fig.~\ref{fig:Calibration}), the image of a non-scanning laser line is used as a common reference point. (The laser line is visible in the image due to the excitation of autofluorescence from the bundle itself). Therefore, the vertical difference in positions between the laser lines in the two images is the offset in pixels. 

The first step in determining the offset was to set the galvo voltage at a value which produces a laser line approximately in the center of the bundle. For both images, intensity profiles are plotted along a line of pixels perpendicular to the laser line. A threshold filter is applied to remove the background, and each profile is then fitted with a Gaussian, and the difference in the center-of-mass positions between the two Gaussian fits is reported as the offset between the images. The line position is updated in real-time to allow instant feedback. The required offset can then be set by manually adjusting the tilting mirror until the desired value is obtained.

The second calibration step, illustrated in Fig.~\ref{fig:Calibration}(d-e), is a fully automated mapping of the galvo drive signal (in volts) to the position of the laser line on the CMOS camera (in pixels). Mapping is done by sweeping through at least 400 voltage values (and hence line positions) and capturing an image at each voltage. If the laser line is visible, an intensity profile is taken across the line to locate the pixel number at the center of the line. A plot is produced of pixel number versus galvo voltage. The slope of a linear fit to this plot gives the scaling between voltage and pixel value, and the intercept gives the offset. These values, combined with the rolling shutter's known progression speed, allow a galvo drive voltage ramp to be generated, which, when triggered by the camera's strobe output trigger, results in the laser line being synchronized with the position of the rolling shutter.

The third step (Fig. \ref{fig:Calibration}(f)) is to locate the two bundle images on the CMOS camera. First, the bundle is pointed towards a light source and the pair of images recorded by the camera. A binary threshold mask is then used to identify each bundle. Each bundle's center is then taken to be the row and column with the largest number of non-zero pixel values. 

As a quality check, we imaged a USAF 1951 resolution target in transmission and compared the confocal and residual images extracted using the calibrated bundle positions. Visually there was some misalignment between fiber cores near one edge, but this did not exceed a distance of 1 core. The effect can be quantified using the mean of the absolute differences between the images, which was found to be 5.5\%. After applying the filter, used to remove the core patterns during imaging, this was further improved to 0.74\%. 

\section{Results and Discussion}

\subsection{Optical Sectioning and Signal Drop}
The optical sectioning performance was characterized by generating profiles of the collected signal as a function of defocus. These profiles were generated by translating the fiber bundle tip using a mechanical translation stage toward a metal plate stained with a fluorescent highlighter. The mean intensity value of the central 250x250 pixel region (240 x 240 \textmu m) of the bundle was extracted for each position of the translation stage.

\begin{figure}
    \centering
    \includegraphics[width=10.1cm]{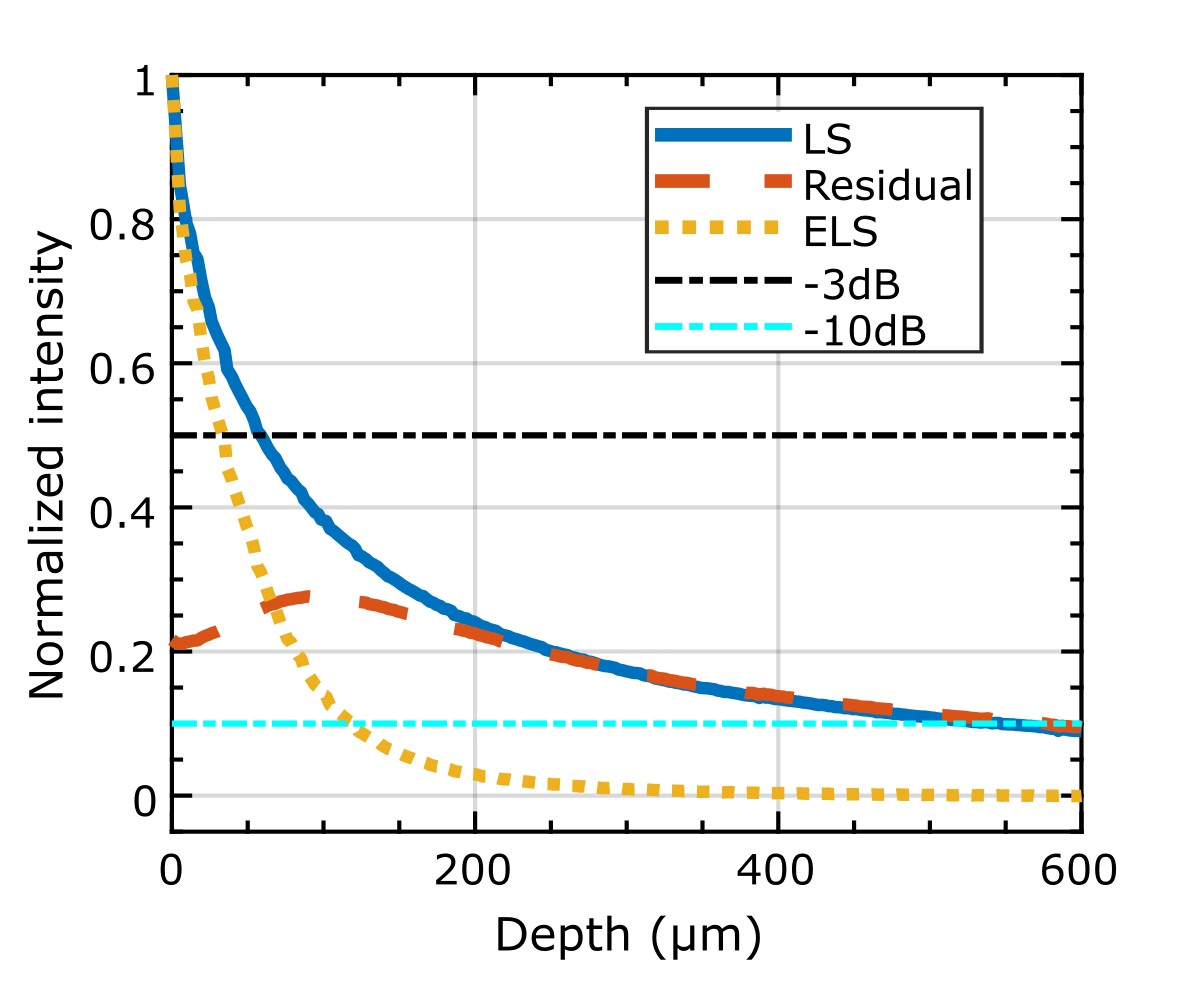}
    \caption{Representative axial response profile for a 2.9~\textmu m slit width and a 29~\textmu m offset. LS and ELS curves were independently normalized, the residual curve was normalised using the same factor as the LS curve for comparison. The 3~dB and 10~dB drop-off distances for LS and ELS can be read-off at the intersection with the respective horizontal lines.}
    \label{fig:repreAxialResp}
\end{figure}

A representative axial profile for a 2.9 \textmu m slit width and a 29 \textmu m slit offset is shown in Fig.~\ref{fig:repreAxialResp}. Plots for both conventional line-scan confocal (LS) and enhanced line-scan (ELS) are shown. Both plots were independently normalized to give a signal of 1 at focus. The 3~dB and 10~dB drop-off distances occur at the half-width half-maximum and half-width tenth-maximum of the LS and ELS curves. Also shown is the residual image profile (acquired with the offset slit), normalized to the same scale as the LS curve. The ELS curve is, therefore, the (re-normalized) result of subtraction of the residual curve from the LS curve. A clear improvement both in the 3~dB and particularly the 10~dB drop-off distances can be seen.

\begin{figure}
    \centering
    \includegraphics[width = 13.5cm]{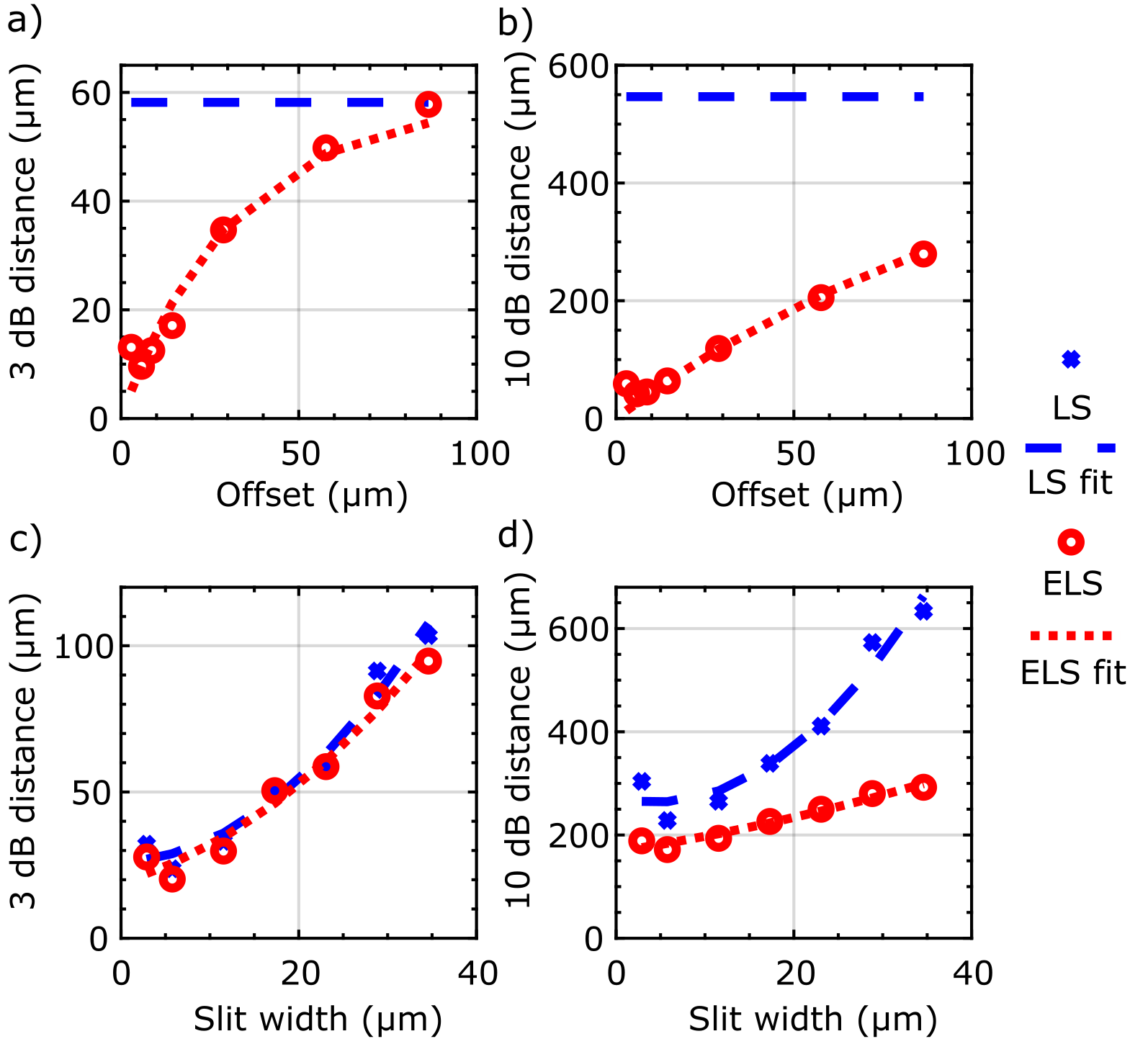}
    \caption{3 and 10~dB drop-off distances as a function of offset and slit width. (a,b) 3~dB and 10~dB drop-off distances for a constant 2.9 \textmu m slit width. (c,d) 3~dB and 10~dB drop-off distances for a constant 77 \textmu m slit offset. The LS reference drop-off distances in (a) and (b) were determined by averaging drop-off distances over 7 runs, with mean 3~dB drop-off distance of $58 \pm 3$ \textmu m and mean 10~dB drop-off distance of $550 \pm 20$ \textmu m.  (a) ELS fit: $58 \; (1-\exp(-0.032 \;\mathrm{OF})$ \textmu m, (b) ELS fit: $550 \;(1-\exp(-0.0084 \;\mathrm{OF}))$ \textmu m, (c) LS fit: $0.065\;\mathrm{SW}^2+0.084\;\mathrm{SW}+26$ \textmu m, ELS fit: $0.042\;\mathrm{SW}^2+0.83\;\mathrm{SW}+19$ \textmu m, (d) LS fit: $0.44\;\mathrm{SW}^2-4.0\;\mathrm{SW}+270$ \textmu m, ELS fit: $0.034\;\mathrm{SW}^2+2.6\;\mathrm{SW}+170$ \textmu m, where $\mathrm{OF}$ is the offset, and $\mathrm{SW}$ is the slit width.}
    \label{fig:axialRespPlotsCombined}
\end{figure}

The change in the optical sectioning as a function of residual image offset (with a constant slit width) is shown in Fig. \ref{fig:axialRespPlotsCombined}(a,b), parameterized by both the 3~dB and 10~dB drop-off distances.  For comparison, the 3~dB and 10~dB drop-off distances are also shown for LS operation (i.e., without subtraction of the residual image, in which case the image offset has no effect). The ELS plots show that a smaller offset leads to greater optical sectioning strength improvement, shown by a decrease in both the 3~dB and 10~dB drop-off distances. As the offset increases, the 3~dB drop-off distance for ELS converges with the LS much faster than the 10~dB drop-off distance, so that even with very large offsets, there remains an optical sectioning improvement far from focus.

The change in the optical sectioning as a function of slit width (with a constant residual image offset) is shown in Fig.~\ref{fig:axialRespPlotsCombined}(b,c). Reducing the slit width improves both the 3~dB and 10~dB drop-off distances for both LS and ELS images, as would be expected. It is also apparent that the 10~dB drop-off distance is much more drastically affected by the slit width for the LS case than the ELS case; this is because ELS is very successful at removing far-from-focus light even when the slit width is large. The 10~dB drop-off improvement shows that, at least in principle, ELS can allow larger slit widths to be used than LS.

The pixelation of the fiber imposes an additional constraint on sectioning strength: once the offset or the slit width approaches the inter-core spacing of the bundle, there are virtually no gains in optical sectioning strength. The laser line at the distal end of the fiber cannot be made significantly smaller than the core spacing, and so reducing the detection slit below this has little effect. Similarly, if the slit separation falls below the inter-core spacing, approximately 3~\textmu m, the offset slit begins primarily sampling in-focus light and so subtraction lowers the signal levels. This effect is compounded by the fact the intensity across the \say{line} has a Gaussian profile, so that the intensity from the Gaussian tail bleeds into other cores, as well as the effects from any cross-core coupling. 

To demonstrate this, axial response profiles with a constant offset of 29~\textmu m and slit widths of 1.9, 2.9 and 5.8 \textmu m were generated; the respective LS 3 dB drop-off distances were 70, 70, and 67 \textmu m, and ELS distances were 120, 120, and 120~\textmu m. Offsets of 2.9, 5.8, and 8.6 \textmu m have respective 3~dB drop-off distances of 13, 9.6, and 12.5 \textmu m and respective 10~dB drop-off distances of 59, 42, and 45~\textmu m, again showing that there is no continued improvement from smaller offsets once the core spacing is approached.

The ELS subtraction approach is not without penalty. Since the residual image does not have exactly zero intensity from in-focus depths, there is a drop in the in-focus signal relative to LS when subtraction is performed. The signal drop occurs primarily because the illumination line is not arbitrarily thin; it has a Gaussian profile such that the region of the bundle sampled by the offset slit will still receive some illumination. Scattering or core leakage within the bundle will also tend to broaden the line further. 

\begin{figure}
    \centering
    \includegraphics[width = 13cm]{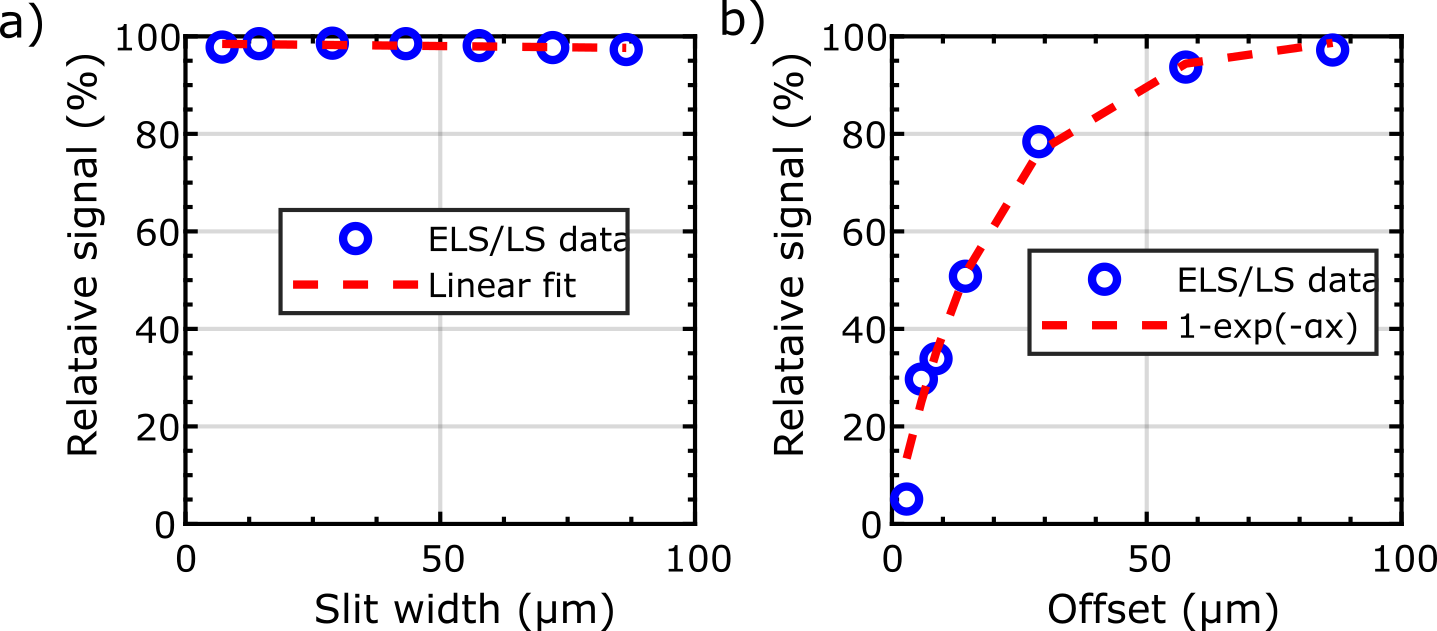}
    \caption{Percentage of in-focus signal maintained (relative to LS) when varying slit width and offset. (a) Shown as a function of slit width with constant 77~\textmu m offset. (b) Shown as a function of slit offset with a constant 2.9~\textmu m slit width. Fit determined through non-linear least squares curve fit $\alpha = 0.05$. Signal reported as relative intensity of ELS to LS when bundle is in contact with mirror. }
    \label{fig:sigdrop}
\end{figure}

The signal level at focus was compared between LS and ELS using the un-normalized values of the axial response profiles when the fiber was in contact with the metal plate (i.e., at zero defocus). The resulting drop of the in-focus signal as a function of slit width and offset are shown in Fig.~\ref{fig:sigdrop}. With constant slit width, it is clear that the smallest in-focus signal drop occurs at the largest offset, as would be expected. At a 77~\textmu m offset, there appears to be virtually no price paid in terms of relative in-focus signal drop when using ELS instead of LS. It may therefore be tempting to choose a large offset to minimize this loss of signal while still obtaining improved sectioning. However, as discussed in Sect.~\ref{discussion} a large offset leads to significant edge artifacts. 

\subsection{Reduction of Motion Artefacts}
\begin{figure}
\centering\includegraphics[width=12cm]{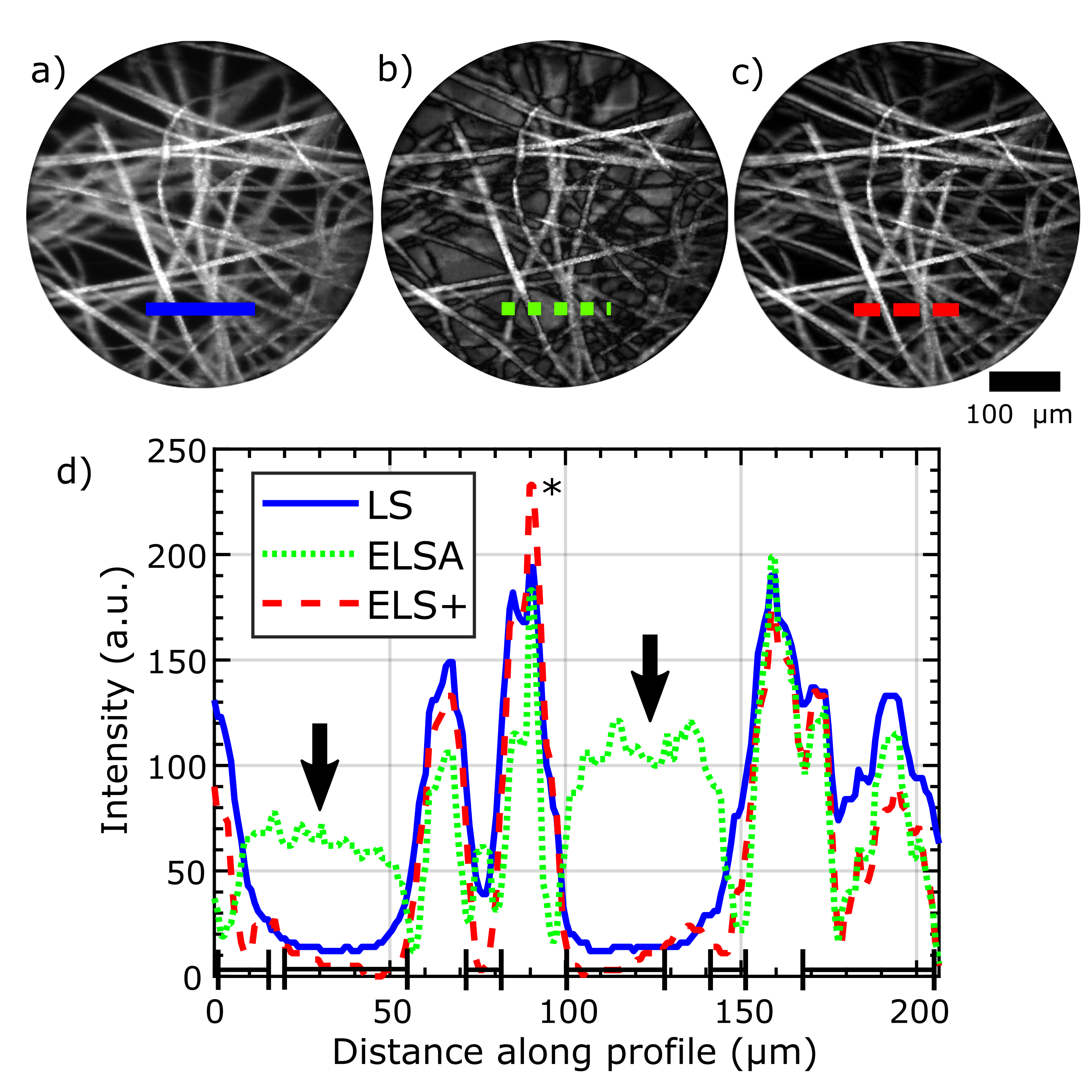}
\caption{Images of lens tissue paper stained with fluorescent highlighter overlaid on a fluorescent background. (a) LS, (b) ELSA, (c) ELS+, and (d) intensity profiles. The probe was moved by hand. The imaging frame rate was set to 10~Hz. In (a) lower contrast is caused by light returning far-from-focus, while in (b) motion artifacts are clearly observed, and the subtraction fails to achieve improved sectioning. In (d), square brackets indicate improved sectioning with ELS+, and arrows indicate some regions where motion artifacts appear in ELSA images. LS and ELS+ images were independently normalized to 0.01\% saturation, which led to the intensity of the ELS+ image being greater than the LS image in the marked region (*). The same factor normalized ELSA images as ELS+ images. Acquired at 10 Hz reconstructed from recording. }
\label{lp}
\end{figure}

The primary advantage of the simultaneous acquisition of the two images in the ELS+ approach reported here, compared to the previously reported ELSA approaches using sequential acquisition, is the reduction of motion artifacts. Motion artifacts are a function of frame rate, and so to show this most clearly, images of fluorescently stained lens tissue paper placed over a fluorescent background were acquired at a frame rate of 10 Hz with a moving probe. Fig. \ref{lp}(a) shows an image acquired using LS only (no subtraction); the presence of background and an out-of-focus signal is evident. In Fig. \ref{lp}(b), the ELSA subtraction (using sequential acquisition) has led to noticeable motion artifacts. In Fig. \ref{lp}(c), the ELS+ image (with the two images acquired simultaneously) shows the removal of the out-of-focus background without the motion artifacts. For comparison, the LS and ELS+ images were individually normalized, allowing 0.01\% saturation, while the same normalization factor scaled the ELSA as ELS+.

\subsection{Edge Effect Artefacts}
\label{discussion}
While the ELS+ approach does not introduce motion artifacts, there are some observed edge artifacts which are shared with the previously reported ELSA approach, and some which are common to most fiber bundle endomicroscopes. Edge artifacts limit the visual quality of mosaics because the same region of tissue is samples with different parts of the bundle; this is particularly apparent when using real-time mosaicing with dead-leaf image insertion rather than blending.

Firstly, all fiber bundle systems that are not fully confocal have inherently better sectioning near the edge of the bundle than at the center for geometrical reasons (simply, less excitation light reaches the out-of-focus depths in the edge areas). A system with better sectioning performance inherently reduces the relative effect of this artifact, so ELS+ is an improvement over conventional LS in this regard.  

\begin{figure}
    \centering
    \includegraphics[width=12cm]{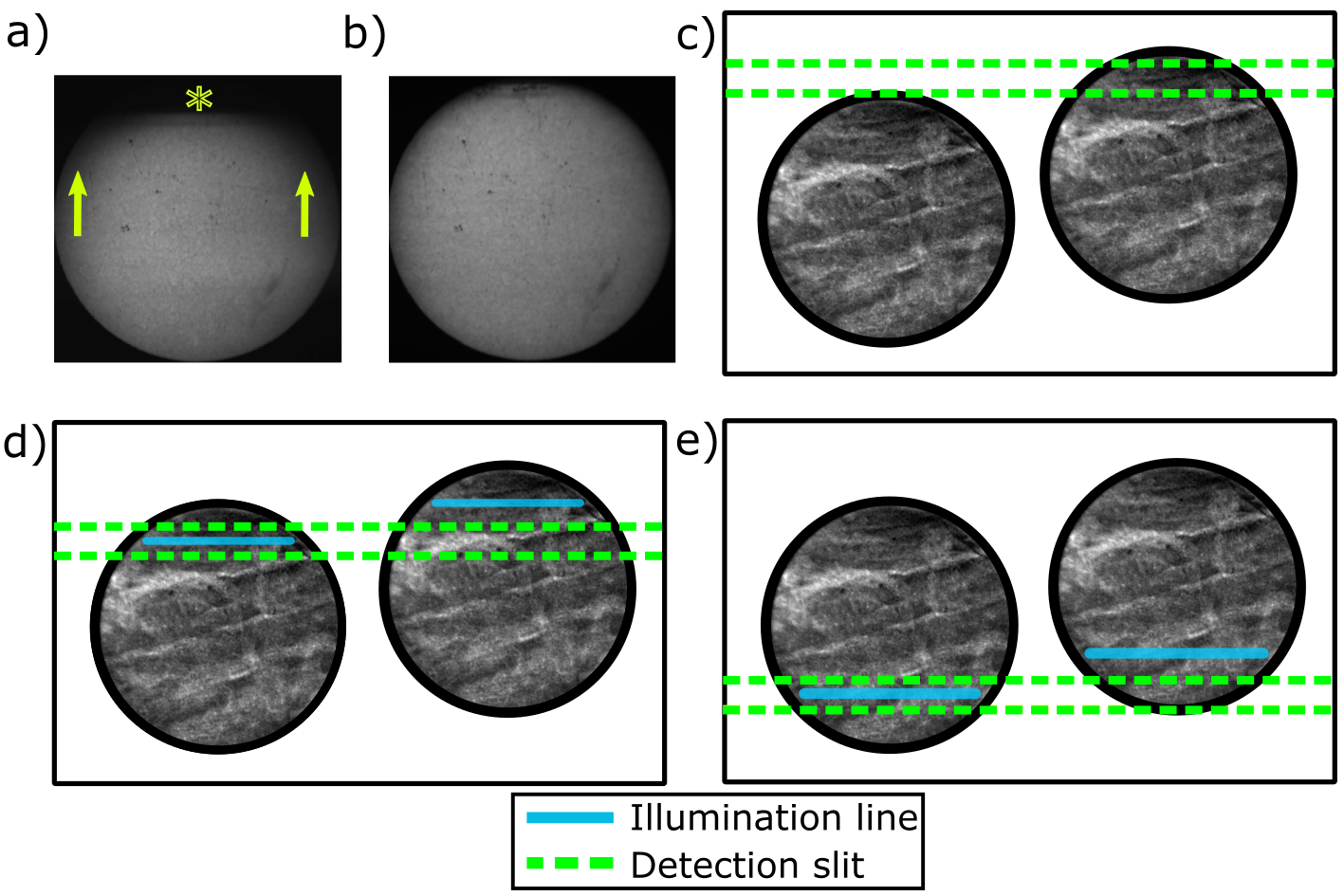}
    \caption{Demonstration of edge effect artifacts in residual images. (a,b) Average of \textasciitilde 300 residual images taken with a moving probe over paper stained with a yellow highlighter, acquired using a constant slit width (8.8 \textmu m) and varying offset: (a) 74 \textmu m (b) 12 \textmu m. The intensity of (a) was scaled to match (b) in the central 100x100 pixel region. Experimental causes of artifacts: (c) illumination line clipping - seen in image (a*), (d) greater detection area in residual images (a-arrow), (e) greater illumination area in residual images - less pronounced in (a). }
    \label{fig:residualartefacts}
\end{figure}

Specifically for ELS (both ELSA and ELS+), there are further edge artifacts, due to the shape of the bundle, which are illustrated in Fig. \ref{fig:residualartefacts}, and which have not been commented on previously. Because the detection slit is always leading the illumination line, when the detection slit is near the top of the bundle, the illumination line is above the bundle, and no light reaches the tissue. Hence, near the top of the bundle, the residual image has close to zero intensity, and subtraction does not improve optical sectioning in this area. This can clearly be seen in the residual images presented in Fig. \ref{fig:residualartefacts}(a), although it is not particularly noticeable in individual ELS+ images.

A similar artifact appears even once the illumination line has reached the bundle. Because the bundle is circular, the width of the line changes as it scans through the tissue. Therefore the left and right parts of the residual images will tend to under-estimate the residual signal in the top half of the bundle and over-estimate it in the bottom part. This is illustrated in Fig.~\ref{fig:residualartefacts}(d,e).

Therefore, while a large slit offset for the residual image appears to deliver good performance in terms of the 10~dB drop-off distance, with minimal loss of in-focus signal,  this comes at the expense of worsening the edge artifacts. Reducing the slit offset results in better sectioning and minimizes artifacts, but also results in a lower signal-to-noise ratio. Fig. \ref{fig:residualartefacts}(b) shows this very clearly; when a smaller offset of 11.53~\textmu m is used, there are virtually no artifacts from clipping or unequal illumination areas compared to when a larger offset of 73.7~\textmu m is used as in Fig. \ref{fig:residualartefacts}(a).

A flat-field correction can compensate for illumination differences but would not correct the clipping artifact since there is no way to sample the residual image in the edge regions. Therefore, clipping artifacts can be considered a fundamental limitation of the ELS technique, although not a serious one in practice providing a careful choice of offset is made.
 
\subsection{Ultra-High-Speed Mosaicing}

\begin{figure}
    \centering
    \includegraphics{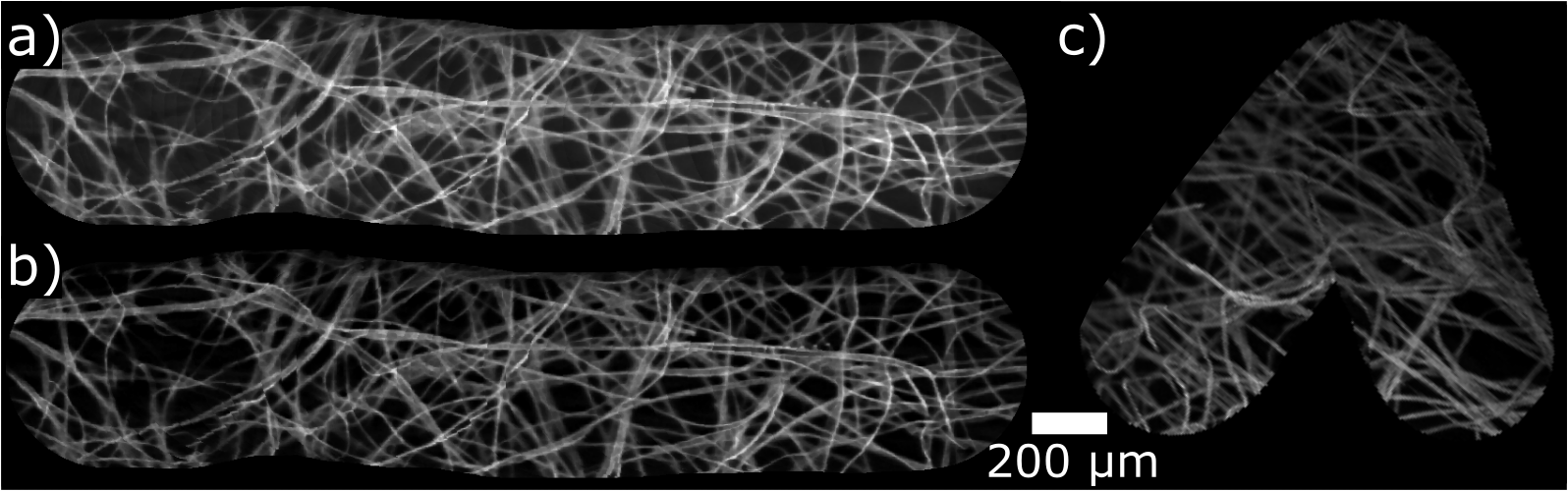}
    \caption{Lens paper stained with yellow highlighter overlaid on a fluorescent background. Acquired at 240 Hz. (a) LS, (b) ELS+ reconstructed from recording, (c) ELS+ real-time mosaicing. 8.8 \textmu m resolution.}
    \label{fig:240hzmosaics}
\end{figure}

Using the system set up in 240~Hz mode, we collected mosaics from lens tissue paper, stained as earlier. Fig~\ref{fig:240hzmosaics}(a) and (b) show mosaics generated from recordings for LS and ELS+, allowing a direct comparison, while Fig~\ref{fig:240hzmosaics}(c) shows an ELS+ mosaic generated and displayed in real-time. To the best of our knowledge this is the highest speed mosaicing reported from fiber bundle endomicroscopy and is an order of magnitude faster than typically reported rates. Therefore, this higher mosaicing speed provides a significant potential improvement in the usability of mosaicing during high-speed motion of the probe.

\subsection{Tissue Imaging}
\begin{figure}
\centering\includegraphics[width=10cm]{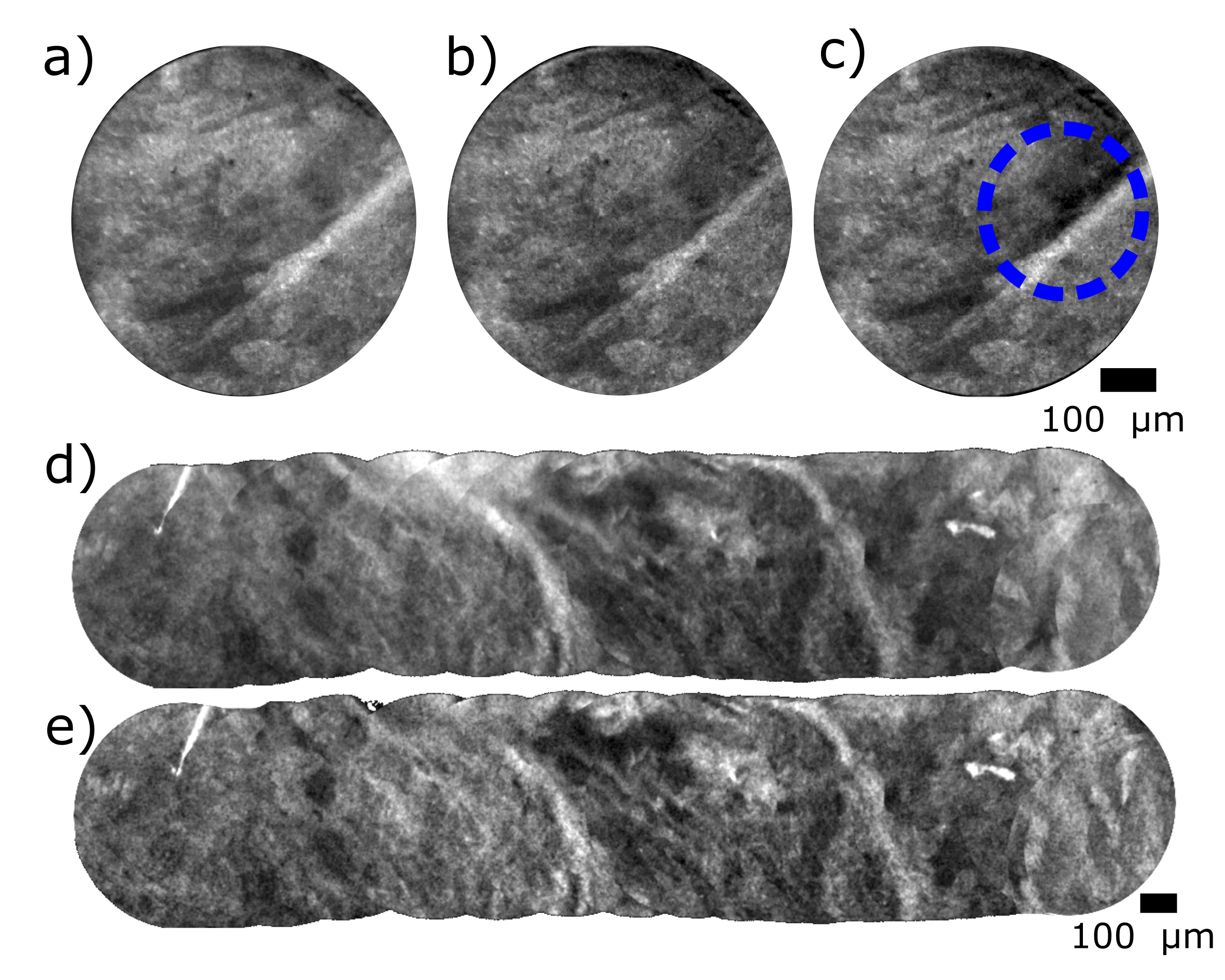}
\caption{Porcine oesophagus labelled with acriflavine hydrochloride. (a) LS, (b) ELS+, (c) ELSA. Circled region indicates light which has been incorrectly subtracted due to a shift between acquisitions. LS and ELS+ independently normalised to 0.01\% saturation, ELSA normalised to same factor as ELS+. Mosaics (d) LS, (e) ELS+. Scale bars 100 \textmu m. 6.8 pixels per core. Acquired at 60 Hz, reconstructed from recording. }
\label{po}
\end{figure}

We validate the ELS+ approach for tissue imaging using \textit{ex vivo} porcine esophagus stained with acriflavine hydrochloride. The confocal image and the residual image were simply extracted from alternating frames to simulate ELSA (sequential acquisition) imaging for comparison. Example individual frames (during motion) are shown in Fig.~\ref{po}(a-c) for ELS, ELSA, and ELS+, respectively. The presence of motion artifacts in the ELSA image, which are not present in the ELS+ image, demonstrates the new approach's value. The improved sectioning of ELS+ over LS can be seen most clearly in the mosaics of  Fig.~\ref{po}(d,e).

\section{Conclusions}

The newly reported approach of simultaneous acquisition enhanced line-scanning endomicroscopy (ELS+) provides improved optical sectioning over line-scanning (LS) endomicroscopy while maintaining the same high frame rate and (most importantly) does not introduce artifacts due to inter-frame motion. Some edge artifacts due to the ELS approach remain, but these can be mitigated through a careful choice of operating parameters, mainly the slit offset. 

The system was validated with topically stained \textit{ex vivo} porcine esophagus and showed improved sectioning compared with LS endomicroscopy and a reduction in motion artifacts compared to sequential acquisition enhanced line-scanning (ELSA). ELS+ has not yet been tested \textit{in vivo} with intravenous fluorescein, so it cannot be confirmed whether the optical sectioning is sufficient in this scenario.

The ELS+ approach also allowed the frame rate to be further improved to 240~Hz using a smaller camera region. This enables mosaicing of optically sectioned images at 240~Hz, faster than previously reported approaches and more than an order of magnitude faster than commercially available systems. 

The approach was tested using a bare fiber bundle imaging probe without any distal optics. In several respects, this underplays the potential advantage of ELS+. Firstly, a bare probe with sharp edges tends to glide over tissue less smoothly than a lensed probe, meaning that the maximum probe speed was as much constrained by this as it was by the need to avoid artifacts. For a lensed system, it would be possible to scan the probe much faster, and hence the removal of motion artifacts would become more critical.

Lensed probes also usually demagnify the fiber bundle onto the tissue, improving the resolution but decreasing the field of view. For example, ELS endomicroscopy's initial demonstration used a probe with a field-of-view of just 240~\textmu m \cite{RN68}. In this case, the higher frame rate becomes more critical for achieving mosaicing to allow sufficient overlap between images, so the benefit of ELS+ will be more pronounced. A high frame rate is also important if multi-frame techniques such as resolution enhancement \cite{RN76} or axial scanning are to be implemented, with the latter also standing to benefit from the improved optical sectioning of ELS+.

\subsection{Acknowledgments}
This work was supported by the Engineering and Physical Sciences Research Council (EPSRC); "Ultrathin fluorescence microscope in a needle" Award No. EP/R019274/1. Andrew Thrapp is supported by a University of Kent Vice Chancellor's Ph.D. Scholarship. 



\bibliography{report}   
\bibliographystyle{spiejour}   

\subsection{Biographies}
\vspace{2ex}\noindent\textbf{Andrew D. Thrapp} is a Ph.D. student at the University of Kent. He received his bachelor's in Physics with a minor in Mathematics from the University of Colorado in 2017. His current research interests include optical-sectioning in microscopy and endomicroscopy. 

\vspace{2ex}\noindent\textbf{Michael R. Hughes} is a Lecturer in Applied Optics in the School of Physical Sciences at the University of Kent. His research interests are in endoscopic and needle-based microscopes and computational approaches to low-cost and point-of-care microscopy.

\end{spacing}
\end{document}